\documentclass[preprint,12pt]{elsarticle}

\usepackage{amsmath}
\usepackage{mathtools}
\usepackage{mdwmath}
\usepackage{mdwtab}
\usepackage{amssymb}
\usepackage{amsthm}

\usepackage{algorithm}
\usepackage{algcompatible}

\usepackage{caption}
\usepackage{subcaption}

\usepackage{float}

\usepackage{epstopdf}
\usepackage{multirow}
\usepackage{color}

\journal{ACM Transactions on Cyber-Physical Systems}

\usepackage{color}

\begin{document}

\begin{frontmatter}

\title{Design and Implementation of Secret Key Agreement for Platoon-based Vehicular Cyber-Physical Systems}

\author[cister]{Kai~Li}
\ead{kaili@isep.ipp.pt}

\author[csiro]{Wei~Ni}

\author[cister]{Yousef~Emami}

\author[csiro]{Yiran~Shen}

\author[cister]{Ricardo~Severino}

\author[cister]{David~Pereira}

\author[cister]{and Eduardo~Tovar}

\address[cister]{Real-Time and Embedded Computing Systems Research Centre (CISTER), Porto, Portugal}

\address[csiro]{Data61, Commonwealth Scientific and Industrial Research Organization (CSIRO), Australia}

\begin{abstract}
In platoon-based vehicular cyber-physical system (PVCPS), a lead vehicle that is responsible for managing the platoon's moving directions and velocity periodically disseminates control messages to the vehicles that follow. Securing wireless transmissions of the messages between the vehicles is critical for privacy and confidentiality of platoon's driving pattern. However, due to the broadcast nature of radio channels, the transmissions are vulnerable to eavesdropping. In this paper, we propose a cooperative secret key agreement (CoopKey) scheme for encrypting/decrypting the control messages, where the vehicles in PVCPS generate a unified secret key based on the quantized fading channel randomness. Channel quantization intervals are optimized by dynamic programming to minimize the mismatch of keys. A platooning testbed is built with autonomous robotic vehicles, where a TelosB wireless node is used for onboard data processing and multi-hop dissemination. Extensive real-world experiments demonstrate that CoopKey achieves significantly low secret bit mismatch rate in a variety of settings. Moreover, the standard NIST test suite is employed to verify randomness of the generated keys, where the p-values of our CoopKey pass all the randomness tests. We also evaluate CoopKey with an extended platoon size via simulations to investigate the effect of system scalability on performance. 
\end{abstract}


\end{frontmatter}

\section{Introduction}
\label{sec_intro}
\begin{figure}[htb]
\centering
\includegraphics[width=5in]{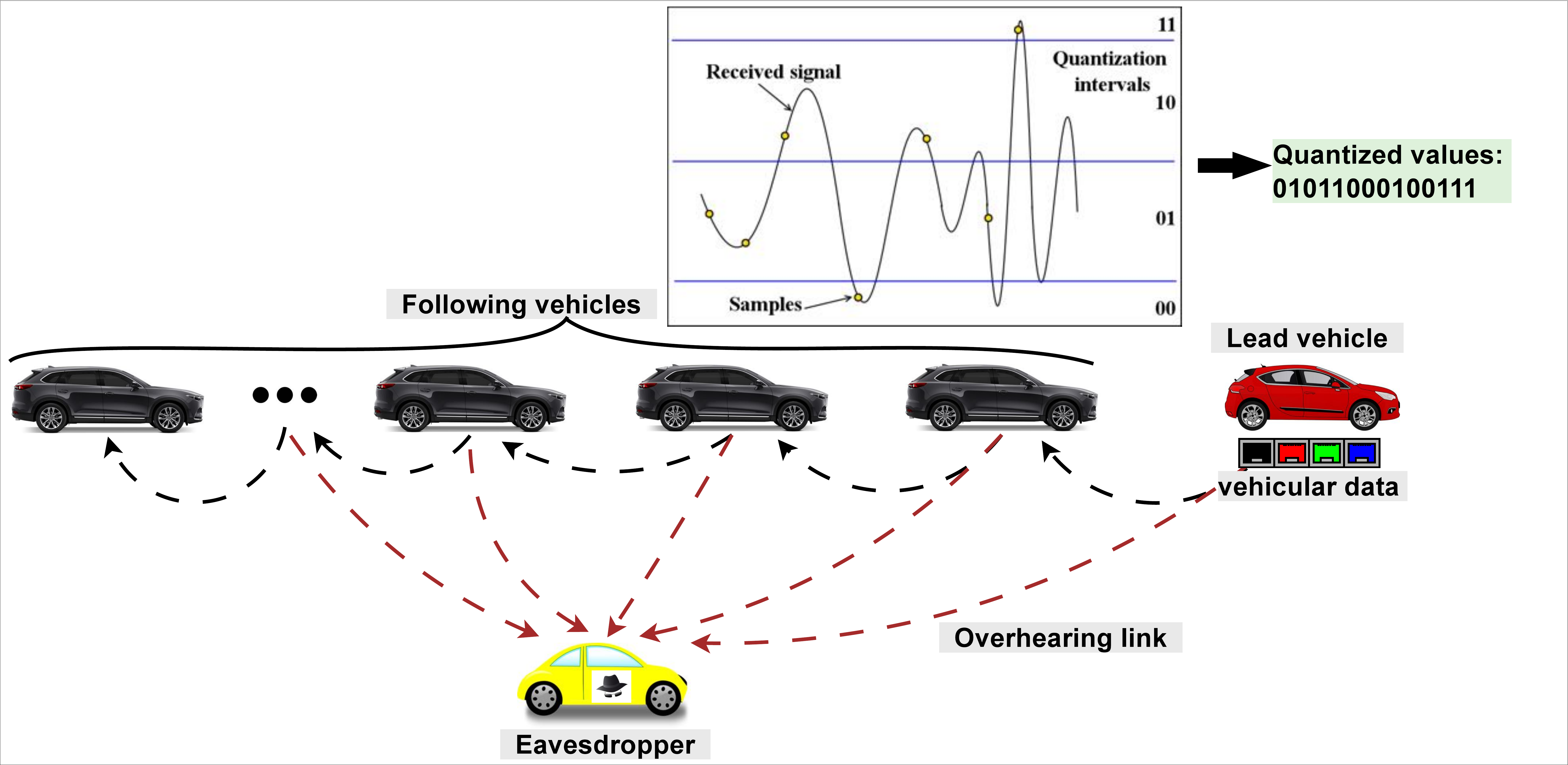}
\caption{In PVCPS, a lead vehicle transmits control messages to update the platoon's driving status. The following vehicles act as data-forwarding nodes for command dissemination, which can be overheard by an eavesdropper vehicle. In this example, a simple method with fixed quantization intervals is used to quantize samples of the received signal. The quantization will output 010110000100111. } 
\label{fig_pvcps}
\end{figure}

In the past few years, advances in autonomous vehicles and inter-vehicle wireless communications have enabled a new platoon-based driving pattern, especially on highways, where the lead vehicle is manually driven and the others follow in a fully automated manner. Vehicular platoon is regarded as a promising driving concept and has been verified to significantly improve road capacity and safety of automated highway systems, and accordingly reduces the traffic congestion (e.g., Safe Road Trains for the Environment Project~\cite{chan2012overview}, SafeCop Project~\cite{pop2016safecop}, and ENABLE-S3 Project~\cite{enables3}). The vehicular platoon can also reduce the fuel consumption and exhaust emissions by 4.7--7.7\% due to air drag reduction between the two vehicles~\cite{al2010experimental}. 
Platoon-based Vehicular Cyber-Physical Systems (PVCPS) are characterized to provide wireless connectivity to vehicular platoons, where the vehicle is equipped with a wireless communication interface on board~\cite{jia2016survey,jia2014network}. For managing the platoon in PVCPS, the lead vehicle controls the platoon's driving status, including driving speed, heading directions, and acceleration/deceleration values, which indicates emergent road conditions, such as traffic jams, crossroads, obstacles or car accidents. As shown in Figure~\ref{fig_pvcps}, the lead vehicle periodically transmits control messages to update the platoon's vehicles with the driving status. The following vehicles in PVCPS act as data-forwarding nodes so that messages from the leader can be disseminated to all vehicles in the platoon~\cite{li2012toward}. In particular, the preceding vehicle disseminates the command to its following vehicle based on store-and-forward broadcasts at different time slots without causing interference to the other vehicles in the platoon~\cite{li2018lcd}. Due to the broadcast nature of radio channels, vehicular command dissemination in PVCPS is vulnerable to eavesdropping attacks~\cite{wasicek2014aspect,shiu2011physical}. With the eavesdropped information, adversaries could track the location of vehicles of interest, and launch spoofing, playback, or impersonation attacks to abuse mobility patterns of the platoon. Consequently, a secret key for message encryption/decryption is crucial to support control message confidentiality, integrity, and sender authentication, which is also critical to the driving safety in PVCPS. 

A common method for establishing a secret key is by using public key cryptography. However, public key cryptography requires a fixed key management infrastructure, which is not applicable to real-time data transmission in mobile wireless environments. Although quantum cryptography~\cite{greenemeier2007election} has started to appear recently, it is prohibitively expensive on the implementation. 

Comparing to various physical layer information of radio channel (such as channel phase), Received signal strength (RSS) can be measured by most of current off-the-shelf wireless devices without any modification, and thus presenting significant cost savings. Generating the secret key with RSS measurements on inter-vehicle radio channel is a promising approach~\cite{wan2018physical,zeng2015physical,wang2011fast}, where two adjacent vehicles in PVCPS extract secret bits from the inherently random spatial and temporal variations of the reciprocal wireless channel between them. Moreover, the properties of the channel are unique to the locations of the platooning vehicles in PVCPS. An eavesdropper misaligned with the platoon measures uncorrelated RSS values, which results in different quantization intervals. Thus, the eavesdropper is not able to generate the same secret key as the platooning vehicles. In addition, RSS varies over time due to motion of the vehicles and multipath propagation. The temporal and spatial variations of RSS can randomize the generated secret key, which enhances security of the RSS-based secret key generation. Particularly, all the vehicles in PVCPS have to agree upon a unanimous secret key so that the disseminated command from the preceding vehicle can be successfully decoded by the following one. However, two critical challenges arise in the secret key agreement. First, the RSS measurements obtained between a pair of vehicles cannot be transmitted over the insecure public channel that is observable to the eavesdropper vehicle, making it hard to reach key agreement for multiple vehicles. Second, previous works on RSS based secret key generation mainly focused on improving the secret bit generation rate between a pair of nodes (by exploiting multiple antenna diversity~\cite{zeng2010exploiting}, temporally and spatially correlated channel coefficients~\cite{chen2011secret}, or opportunistic beamforming and frequency diversity~\cite{huang2013fast}). The unanimity problem of key generation over multiple vehicles remains unsolved. 

In this paper, we propose a cooperative secret key agreement (CoopKey) scheme to address both of the above challenges for secure command dissemination in PVCPS. Unlike existing key generations for point-to-point communication, CoopKey focuses on the unanimous secret key generation over multiple nodes, which is used for encrypting/decrypting the command. 
One dissemination cycle consists of two stages, i.e., cooperative secret key agreement (CSKA) followed by encrypted vehicular command dissemination (EVCD), and the two stages interchange periodically until all control commands from the lead vehicle are disseminated to the tail vehicle. 
During CSKA, the vehicles share channel randomness information by transmitting beacon packets. At the end of CSKA, CoopKey cooperatively quantizes the measured/estimated RSS readings on each vehicle in PVCPS. RSS quantization intervals are recursively adjusted until a unanimous secret key can be generated in EVCD. 
Note that relative mobility of the platooning vehicles is low and stabilized by applying an efficient cruise control in PVCPS, e.g., the techniques in~\cite{li2018lcd} or~\cite{xiao2011practical}, which sustains the reliable RSS measurement. 

To evaluate performance and effectiveness of CoopKey in practical environments, a multi-hop command dissemination testbed is built by forming a platoon of Autonomous Robotic Vehicles (ARVs). For onboard data processing and disseminating, a TelosB wireless node that is equipped with an IEEE 812.15.4-compliant RF transceiver is placed on top of the ARV. Motion of the ARV captures random spatial and temporal variations of the reciprocal inter-ARV channel, which is critical to measure the RSS of fading changes. Experiments are conducted along a walking path in front of the building of CISTER Research Centre in Porto, Portugal. The experiments are designed to show the effect of inter-ARV distances, number of RSS quantization intervals, and secret key length on CoopKey. 
The experimental results confirm the feasibility of using CoopKey for multiple vehicles in real-world PVCPS. The results also demonstrate that CoopKey achieves a lower bit mismatch rate (BMMR) than existing non-cooperative key generation schemes. By applying CoopKey, BMMR of the secret key generated by an eavesdropper is higher than 73\%, indicating that any eavesdropper experiencing independent channel fading is not able to obtain the same key as the ARVs. Furthermore, the generated secret key bit streams also pass the randomness tests of the NIST test suite~\cite{rukhin2001statistical}, which validates the effectiveness of CoopKey. 
CoopKey is also evaluated in simulations with an extended platoon size and inter-vehicle distance to study the scalability. 

It is worth mentioning that some preliminary simulation results were reported in~\cite{li2018cooperative}, where three static sensor nodes cooperatively generate a secret key to encrypt/decrypt the transmitted data in a 2-hop wireless sensor network. However, the results in~\cite{li2018cooperative} can be hardly applied to mobile ARVs in PVCPS, since a fixed inter-node distance results in a stable communication connectivity and small RSS variation. Moreover, the static placement of the nodes in~\cite{li2018cooperative} cannot capture the ARV movements on the link characteristics. 

The rest of the paper is organized as follows. Section~\ref{sec_system} presents the command dissemination security protocol. In Section~\ref{sec_cska}, the steps that incorporate CoopKey for secret key agreement are investigated. Implementation of CoopKey is investigated in Section~\ref{sec_implement}. In Section~\ref{sec_exp}, CoopKey is firstly evaluated on a multi-hop autonomous mobile robotics testbed. Next, the scalability of CoopKey in PVCPS is studied via extensive simulations. Section~\ref{sec_relatedwork} presents related work on link-based secret key generation and vehicle network security, followed by conclusions in Section~\ref{sec_cond}.

\section{Communication Protocol and System Model}
\label{sec_system}
In this section, we present a 2-stage command dissemination protocol for the secure data dissemination in PVCPS, followed by a system model.

\subsection{Command dissemination protocol}
\begin{figure}[htb]
\centering
\includegraphics[width=4in]{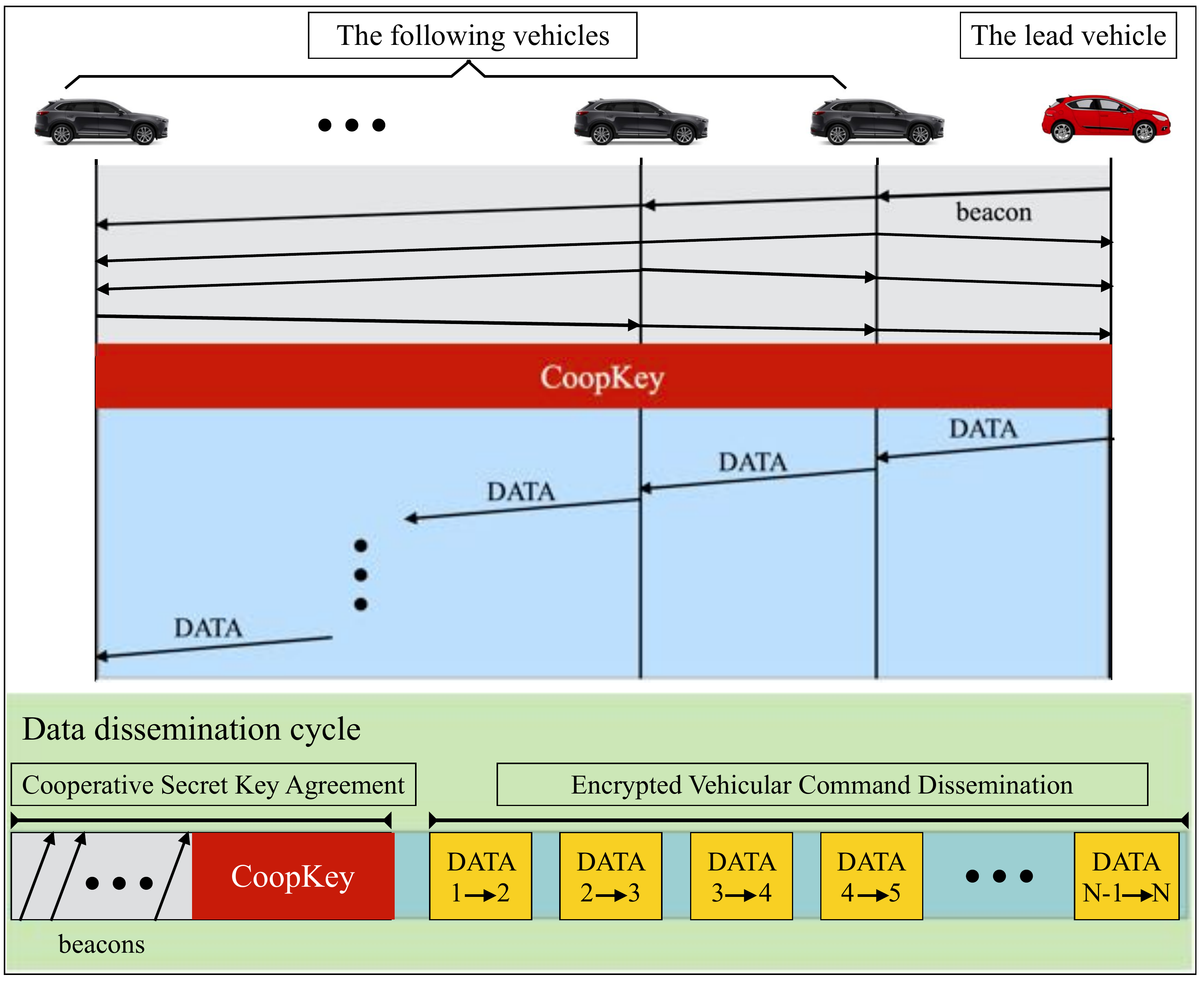}
\caption{The communication protocol for the secret key agreement and command dissemination in PVCPS.} 
\label{fig_protocol}
\end{figure}

Figure~\ref{fig_protocol} demonstrates the 2-stage command dissemination protocol, with CSKA followed by EVCD, where CoopKey is applied for encrypting/decrypting the control command. 
The purpose of CSKA is to share link information among the vehicles, where the vehicles broadcast a single beacon packet in turn. Specifically, the ID number of vehicles fits in the beacon. Transmitting the beacon packet is initialized by the lead vehicle, which solely decides the driving status. Similarly, the adjacent following vehicle broadcasts its beacon packet once the beacon from the lead vehicle is successfully received. Ideally, both the lead vehicle and its adjacent following vehicle should measure the RSS values at the same time by receiving the beacon packets. However, typical commercial wireless transceivers are half duplex, i.e., they cannot both transmit and receive the signals simultaneously. Thus, the two vehicles must measure the radio channel in one direction at a time. Since the time between transmissions of the beacons is much smaller than the inverse of the rate of change of the channel, the measurements have similar RSS readings~\cite{croft2010robust}. 
Note that the beacon packet of the next vehicle can be an acknowledgement to the reception of the preceding vehicle's beacon. In other words, if the beacon of the next vehicle is not received, the preceding vehicle will have to retransmit its beacon packet. 

At the end of CSKA, when all the vehicles in PVCPS finish the beacon transmission, the following vehicles estimate the RSS values between the first two vehicles in the platoon. Next, the CoopKey scheme is carried out to generate a unanimous secret key based on the estimation of the RSS values. Details will be discussed in the next section. 

Note that multiple beacon transmissions can be initialized by the lead vehicle in one dissemination cycle. In this case, CoopKey can be conducted in $Z$ iterations ($Z \geq 1$), and $Z$ secret keys are generated at each vehicle. The larger value of $Z$, the lower RSS estimation errors. This, in turn, leads to a higher likelihood that the generated keys can be unified due to $1 - (1/R_{BMM})^Z$, where $R_{BMM}$ denotes secret bits mismatch rate and $0 \leq 1/R_{BMM} \leq 1$. 

In terms of overhead, consider a 10-vehicle platoon and the lead vehicle transmits 5 beacon packets in one dissemination cycle. The length of a beacon packet is 4 bits. Thus, the total overhead consists of 200 bits, which is much smaller than the size of a data packet. Therefore, the overhead of beacon transmission is negligible due to the small amount of payload. 

For encrypted transmission of the control commands, at the first time slot of EVCD in the dissemination cycle, the lead vehicle uses its secret key generated by CoopKey to encrypt its data packet, and immediately forwards to its next following vehicle. The following vehicles in PVCPS forward the received data packet all the way to the tail vehicle while using their own secret key for the packet decryption. In addition, to enhance the transmission reliability of the data packet, the following vehicles utilize one-hop point-to-point communication in EVCD, which can be supported by both low and high data rate transmissions such as IEEE 802.15.4 in wireless sensor networks~\cite{specificationv1}, or Dedicated Short-Range Communication (DSRC)/ITS-G5 in vehicle networks~\cite{morgan2010notes}. 

\subsection{System model}
As the platoon size is predetermined before forming the platoon, we consider an $N$-vehicle PVCPS, and the command dissemination forms ($N-1$) wireless hops. For the sake of driving safety, the non-leading vehicle in the platoon is required to maintain a certain distance with the preceding one at any time slot $t$, which is denoted by $d_{i,j}(t)$ ($i, j \in [1,N]$). $t \leq T$, where $T$ is the total number of time slots in CSKA. 
Without loss of generality, we consider that the vehicles are traveling with no need to change the platoon size or perform maneuvers (split, merge, leave, etc.), which keeps the operations of cruise control simple. 
In particular, Line of Sight (LOS) communication between the vehicles is available as the antenna can be installed on top of the vehicle, and the platoon travels on the same road segment. Thus, large-scale path loss is considered to model the inter-vehicle communication channel. 

Let $P^{tx}_i(t)$ denote the transmit power (in dB) of a beacon packet at $v_i$. The receive power at $v_j$ that depends on the distance between $v_i$ and $v_j$ can be given by 
\begin{align}
P^{rx}_j(t) = P^{tx}_i(t) + \vartheta - 10 \eta_{PL} \log_{10}(d_{i,j}(t)) + \phi_{i,j}(t),
\label{eq_fading}
\end{align}
where $\vartheta$ is a positive fixed constant relating to the channel, and $\eta_{PL}$ is the path loss exponent. The term $\phi_{i,j}(t)$ denotes the lognormal shadow fading over slot $t$. Thus, we know 
\begin{align}
d_{i,j}(t) = 10^{\frac{H_{i,j}(t) + \vartheta + \phi_{i,j}(t)}{10\eta_{PL}}},
\label{eq_distance}
\end{align}
where $H_{i,j}(t) = P^{tx}_i(t) - P^{rx}_j(t)$ presents the RSS of the channel between sender $v_i$ and receiver $v_j$. 
According to the reception of the beacon packet, the RSS value between $v_i$ and $v_j$ ($i \in [1,2], j \in [3,N]$), i.e., $H_{1,j}(t)$ and $H_{2,j}(t)$, can be measured by the following vehicle $v_j$. 

Given the driving pattern of the platoon, the distance between $v_1$ and $v_2$ can be obtained by $d_{1,2}(t) = d_{1,j}(t) - d_{2,j}(t)$. According to~\eqref{eq_distance}, the RSS of the link between $v_1$ and $v_2$ can be estimated by the other (following) vehicle $v_j$ ($j \in [3,N]$) based on the difference between $H_{1,j}(t)$ and $H_{2,j}(t)$. $10^{\frac{H^j_{1,2}(t) + \vartheta + \phi_{1,2}(t)}{10\eta_{PL}}} = 10^{\frac{H_{1,j}(t) + \vartheta + \phi_{1,j}(t)}{10\eta_{PL}}} - 10^{\frac{H_{2,j}(t) + \vartheta + \phi_{2,j}(t)}{10\eta_{PL}}}$, where $H^j_{1,2}(t)$ denotes the estimate of $H_{1,2}(t)$ at $v_j$. 
Therefore, all the vehicles in PVCPS can have (either measured or estimated) RSS information of the channel between $v_1$ and $v_2$. A unified secret key can be generated at the vehicles in PVCPS once $H_{1,2}(t)$ and $H^j_{1,2}(t)$ are properly quantized by $v_i$ and $v_j$ ($i \in [1,2], j \in [3,N]$), respectively. 
Notations used in the paper are summarized in Table~\ref{tb_variables}. 

\begin{table}[htb]
    \centering
    \caption{The list of fundamental variables}
    \begin{tabular} {|p{1.8cm}|p{5.5cm}|} \hline
        \bf{Notation} & \bf{Definition} \\ \hline
        	$N$ & number of vehicles in PVCPS \\[1pt] \hline
	$v_i$ & the $i$th vehicle in PVCPS \\[1pt] \hline 
	$Z$ &  number of iterations of the key agreement \\[1pt] \hline 
	$T$ & total number of time slots in CSKA \\[1pt] \hline 
	$d_{i,j}$ &  distance between the $v_i$ and $v_j$ \\[1pt] \hline 
	$\vartheta$ &  positive fixed constant relating to the channel \\[1pt] \hline 
	$\eta_{PL}$ &  path loss exponent \\[1pt] \hline 
	$\phi_{i,j}(t)$ &  lognormal shadow fading over time slot $t$ \\[1pt] \hline 
	$P^{tx}_i(t)$ & transmit power (in dB) of a beacon packet at $v_i$ \\[1pt] \hline 
	$H_{i,j}(t)$ &  RSS of the channel between $v_i$ and $v_j$ \\[1pt] \hline 	
	$H^j_{1,2}(t)$ &  estimate of $H_{1,2}(t)$ at vehicle $v_j$ ($j \in [3,N]$) \\[1pt] \hline
	$L$ & total number of quantization intervals \\[1pt] \hline 
	$R^{BMM}_l$ &  number of secret bits that mismatch at the $l$-th quantization interval \\[1pt] \hline 
	$\xi_l^+$ and $\xi_l^-$  & upper and lower thresholds of the $l$-th quantization interval, respectively \\[1pt] \hline 
	$K_i$ & the secret key generated by $v_i$ \\[1pt] \hline 
	$Q$ & length of the Gray codeword \\[1pt] \hline 
   \end{tabular}
\label{tb_variables}
\end{table}

\section{Cooperative Secret Key Agreement}
\label{sec_cska}
In this section, we investigate adaptive RSS quantization and secret key extraction to incorporate CoopKey during CSKA. In addition, an eavesdropper that can also quantize the RSS measurement in attempt to recover the secret key is also discussed. 

\subsection{Adaptive RSS quantization}
Since $H_{1,2}(t)$ measured by vehicles $v_1$ and $v_2$ or $H^i_{1,2}(t)$ estimated by vehicle $v_i$ ($i \in [3,N]$) can be different, due to the motion of the vehicles and multipath fading, the generated secret key bits at the vehicles in PVCPS can be possibly inconsistent if the quantization intervals are not properly determined. In this step, the variations of the RSS are optimally quantized for generating the secret keys. Specifically, $v_1$ and $v_2$ quantize $H_{1,2}(t)$, while the other following vehicles $v_i$ ($i \in [3, N]$) quantize the $H^i_{1,2}(t)$ so that the fading channel randomness can be converted into bit vectors. 
We define $R^{BMM}_l$ as the number of secret bits that mismatch at the $l$-th quantization interval, which yields  
\begin{align}
R^{BMM}_l = \veebar(f_{qnt}(H_{1,2}(t)),f_{qnt}(H^3_{1,2}(t))) + \sum^{N-1}_{j=3} \veebar(f_{qnt}(H^j_{1,2}(t)), f_{qnt}(H^{j+1}_{1,2}(t))), 
\label{eq_def_Delta}
\end{align}
where $0 \leq 1/R_{BMM} \leq 1$, $\veebar(\cdot)$ stands for the operation of XOR, and $f_{qnt}(\cdot)$ is a quantizer to convert RSS measurements into key bits. In particular, $f_{qnt}(\cdot)$ can be given by~\cite{ren2011secret} 

\begin{align}
f_{qnt}(x_i(t)) = \left\{
\begin{array}{l}
1,~\text{if}~\xi_l^- \leq x_i(t) < \xi_l^+;\\
0,~\text{otherwise}. 
\end{array}
\right.
\label{eq_def_qnt}
\end{align}
where $1 < l \leq L$, and $L$ denotes the total number of quantization intervals. $\xi_l^+$ and $\xi_l^-$ denote the upper and lower thresholds of the $l$-th quantization interval, respectively. $x_i(t)$ is the RSS measurement at vehicle $v_i$ in time slot $t$. 

Due to the fact that $\xi_{l-1}^+ = \xi_l^-$, the problem of deriving $\xi_l^-$ for minimizing $R^{BMM}_l$ now is to obtain $\xi_{l-1}^+$, where $l \in (1,L]$. 
Therefore, $\xi_l^+$ and $\xi_l^-$ ($l \in (1,L]$) can be recursively adapted with the aim of minimizing $R^{BMM}_l$. Note that $\xi_1^-$ is the minimum required RSS for decoding the packet, which is known apriori. 
Motivated by this, we propose a dynamic programming approach for achieving a feasible RSS quantization intervals allocation with a polynomial complexity. Specifically, we define the subproblem for the first $l$ intervals by $\Phi_l$, which leads to the minimum mismatch rate of RSS quantizations, as given by 
\begin{align}
\Phi_l = \min_{l^\prime \in (1,l]} \Big\{R^{BMM}_{l^\prime}~|~\xi_{l^\prime}^- < \xi_{l^\prime}^+, \xi_{l^\prime-1}^+ \leq \xi_{l^\prime}^- \Big\}, 
\label{eq_opt_phi}
\end{align}
According to Bellman Equation~\cite{bertsekas2005dynamic}, $\Phi_l$ can be solved recursively, based on the results of all preceding subproblems $\Phi_{l-1}$. It can be given by $\Phi_l = \min \{ \Phi_{l-1}, R^{BMM}_l\}$.

The number of subproblems $\Phi_l$ depends on the total number of quantization intervals, $L$. After solving all the subproblems, the quantization intervals can be given by 
\begin{align}
\{\xi_1^-, \xi_1^+, ..., \xi_L^-, \xi_L^+\} = \arg\min_{l \in (1,L]} \sum^l_{l^\prime=1} R^{BMM}_l
\label{eq_opt_optIntvl}
\end{align}
Backward induction has been widely used to solve dynamic programming problems and can determine a sequence of optimal actions by reasoning backwards~\cite{cormen2009introduction}. It starts by first assessing the last bound of the quantization intervals, i.e., $\xi_L^+$, and then uses the outcome to determine the second-to-last bound, i.e., $\xi_L^-$. This continues until the bounds are decided for all the quantization intervals. The details are presented in Algorithm~\ref{alg_dpa}. 

In terms of time complexity of Algorithm~\ref{alg_dpa}, the number of subproblems to be solved depends on the total number of quantization intervals, $L$. The time complexity of solving each subproblem using~\eqref{eq_opt_phi} is $\mathcal{O}(1)$. The time complexity of backward induction is $\mathcal{O}(L)$. Therefore, the overall time complexity of CoopKey is $\mathcal{O}(L^2)$, which is applicable to a practical PVCPS. 

\begin{algorithm}[htb]
\begin{algorithmic}[1]
\caption{Dynamic programming algorithm with backward induction}
\label{alg_dpa}
\STATE{\textbf{Initialize: } $R^{BMM}_l$, $\xi_1^-$}
\FOR{Each quantization interval $l = 1$ to $L$}
\STATE{Solve $\Phi_l = \min \{ \Phi_{l-1}, R^{BMM}_l\}$ according to~\eqref{eq_opt_phi}.}
\STATE{Record $\{\xi_1^-, \xi_1^+, ..., \xi_l^-, \xi_l^+\}$.}
\ENDFOR
\STATE{\textbf{Backward induction}}
\STATE{$l \to L$.}
\FOR{$l \geq 2$}
\STATE{$\Phi_l \gets R^{BMM}_l$.}
\STATE{Upper threshold: $\xi_l^+$ $\gets$~\eqref{eq_opt_optIntvl}.}
\STATE{Trace backward: $\xi_l^-$ $\to$ $\xi_{l-1}^+$.}
\STATE{$l \to l - 1$.}
\ENDFOR
\STATE{RSS measurement of the vehicle at $t$ is quantized according to~\eqref{eq_def_qnt}.}
\end{algorithmic}
\end{algorithm}

\subsection{Secret key extraction}
After the RSS quantization, an encoding scheme, i.e., $f^{v_i}_\text{encoding}(\xi_l^-, \xi_l^+)$ is utilized to assign a binary codeword to each quantization bin $[\xi_l^-, \xi_l^+]$ for extracting the secret key $K_i$. Specifically, we implement Gray coding as an example of $f^{v_i}_\text{encoding}(\xi_l^-, \xi_l^+)$ as follows~\cite{liu2013fast,ye2006extracting}. 
\begin{itemize}
\item Let $k_i(l),~l \in (1, L]$ denote the complement bit of the codeword, where 
\begin{align}
k_i(l) = \left\{
\begin{array}{l}
1,~l\mod4 \geq 2;\\
0,~\text{otherwise}.
\end{array}
\right.
\label{eq_graybit}
\end{align}
\item Generate a Gray codeword list whose two neighboring codewords only have one-bit difference. Moreover, the list contains $2^{Q}$ possible codewords, where $Q$ denotes length of the Gray codeword. 
\item Define $f^+_i(l) = \lfloor (l-1)/4 \rfloor$. Thus, $K^+_i(l) \in \{0,1\}^Q$ is the $f^+_i(l)$-th Gray codeword. 
\item Define $f^-_i(l) = \lfloor ((l+1) \mod L)/4 \rfloor$. Thus, $K^-_i(l) \in \{0,1\}^Q$ is the $f^-_i(l)$-th Gray codeword. Moreover, $K^-_i(l)$ can be the codeword list that circularly shifts $K^+_i(l)$ by two elements. 
\end{itemize}
Note that $f^{v_i}_\text{encoding}(\xi_l^-, \xi_l^+)$ can be employed by other existing encoding schemes, e.g., Gillham coding, and Lucal coding. 
Based on the codeword of $f^{v_i}_\text{encoding}(\xi_l^-, \xi_l^+)$, well-studied symmetric secret keys can be straightforwardly generated to encrypt and protect the transmissions at every hop. 
Algorithm~\ref{alg_coopkey} depicts the algorithm flow of cooperative secret key agreement in CoopKey. 

\begin{algorithm}[htb]
\begin{algorithmic}[1]
\caption{Algorithm flow of CoopKey.}
\label{alg_coopkey}
\STATE{\textbf{Initialize: } beacons, $N$, $L$, $T$, $Z$.}
\STATE{\textbf{RSS measurement and quantization: }}
\WHILE{Iterations are smaller than $Z$}
\STATE{Beacon packets are broadcasted by the vehicles.}
\STATE{$R^{BMM}_l$ $\gets$~\eqref{eq_def_Delta}.}
\STATE{$\{\xi_1^-, \xi_1^+, ..., \xi_L^-, \xi_L^+\}$ $\gets$ Alg.~\ref{alg_dpa}.}
\ENDWHILE
\STATE{\textbf{Secret key extraction: }}
\FOR{$l \leq L$}
\IF{$v_i \in \{v_1,v_2\}$ $\&$ $f_{qnt}(H_{1,2}(t)) \in [\xi_l^-, \xi_l^+]$}
\STATE{$K_i$ $\gets f^{v_i}_\text{encoding}(\xi_l^-, \xi_l^+)$.}
\ENDIF
\IF{$v_i \in \{v_3,v_4,...,v_N\}$ $\&$ $f_{qnt}(H^i_{1,2}(t)) \in [\xi_l^-, \xi_l^+]$}
\STATE{$K_i$ $\gets f^{v_i}_\text{encoding}(\xi_l^-, \xi_l^+)$.}
\ENDIF
\ENDFOR
\STATE{\textbf{Output: } the $Q$-bit secret key $K_i$.}
\STATE{The secret key $K_i$ is used by $v_i$ $(i \in [1,N])$ to encrypt/decrypt the data.}
\end{algorithmic}
\end{algorithm}

Also note that the proposed CoopKey algorithm is compatible with state-of-the-art secret key reconciliation schemes, such as Cascade~\cite{wei2013adaptive}, low density parity check~\cite{liu2012exploiting} and Turbo code~\cite{epiphaniou2018nonreciprocity}, where the secret bit discrepancies of the key agreement resulting from random channel noises are reconciled for all the vehicles. The overhead of the reconciliation can be reduced by taking advantage of a high key agreement rate achieved by CoopKey. Therefore, CoopKey guarantees the key agreement and correctness at the vehicles in the presence of the RSS measurement randomness.

\subsection{The eavesdropper}
An eavesdropper is typically wavelengths away from the platoon, and can experience an independent radio channel~\cite{durgin2003space}. This is because the eavesdropper can be noticed or detected when it is too close to the platoon (e.g., less than a few wavelengths from the platoon). As the vehicles of the platoon drive at a highway speed in a fully automatic fashion, a dedicated lane is likely to be reserved on the highway for vehicular platoons for driving safety. Any other vehicles taking the reserved lane can be regarded as eavesdroppers. However, an eavesdropper can travel in parallel to a platoon at the similar velocity, while keeping some distance to not be noticed. 

The eavesdropper can overhear the beacon packets during CSKA, and quantize the channels from the platooning vehicles in attempt to recover the secret key. The eavesdropper that attempts to decode cruise control information of the platoon is not interested in disrupting the key agreement in PVCPS. Moreover, the eavesdropper is not able to possess the aprior knowledge of RSS measurements between two arbitrary locations that the platooning vehicles are, since such environmental sensitive information requires significant effort to obtain, e.g., recording RSS fingerprints of every movement along the highway in advance.

\section{Implementation of CoopKey Testbed} 
\label{sec_implement}
\begin{figure*}[htb]
\centering
\includegraphics[width=5.5in]{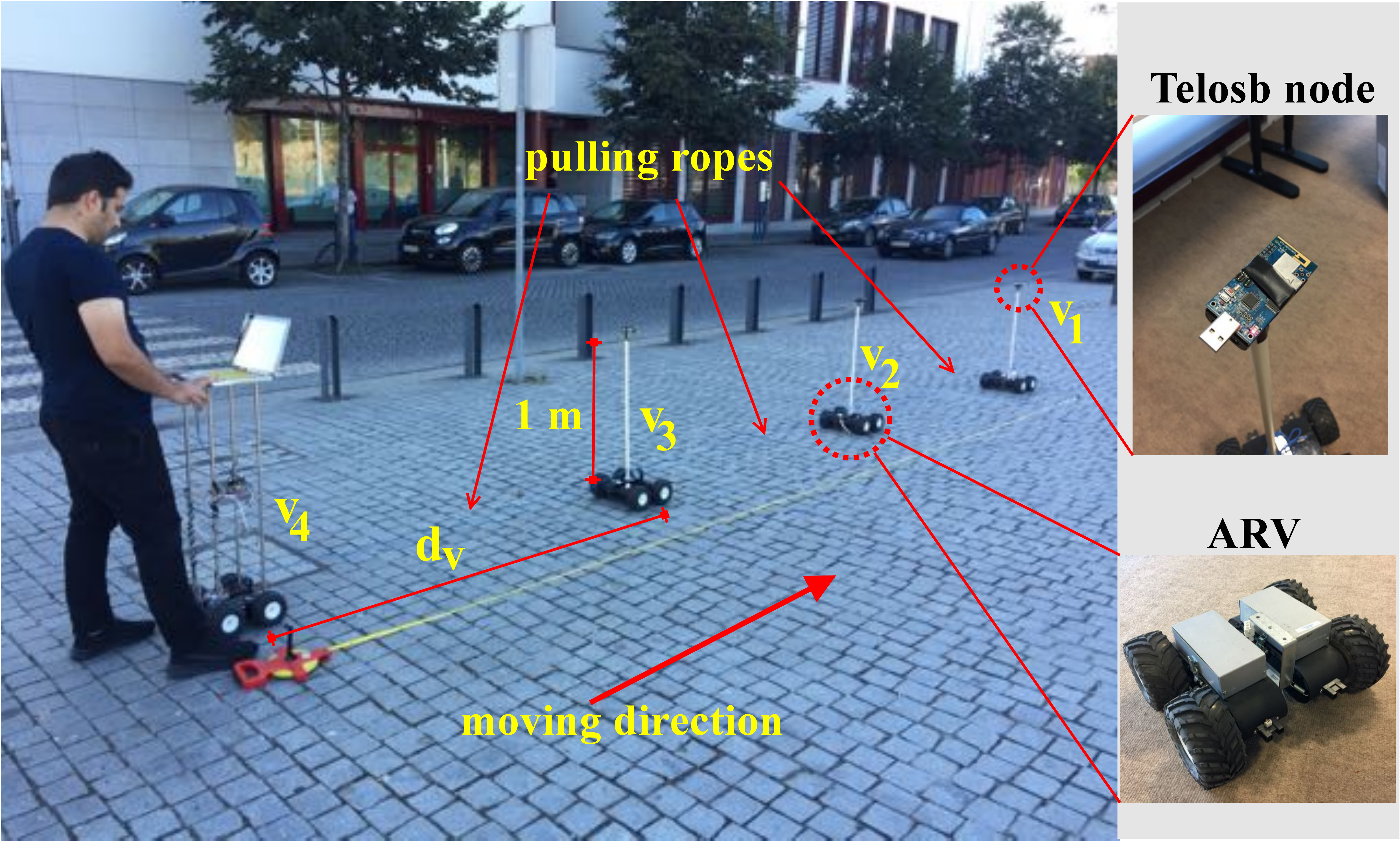}
\caption{The CoopKey testbed is built with 4 ARVs, from $v_1$ to $v_4$, which are physically connected by pulling ropes. The inter-ARV distance is $d_v$. The TelosB node mounted on a 1m-high plastic pole is placed on top of the ARV. The TelosB node at $v_4$ is connected to a laptop via a USB connection for data logging.} 
\label{fig_testbed}
\end{figure*}

We implement CoopKey with the command dissemination protocol on our multi-hop ARV testbed, as shown in Figure~\ref{fig_testbed}. The testbed is built with a platoon of 4 ARVs, from $v_1$ (the lead ARV) to $v_4$ (the tail ARV). Particularly, the two adjacent ARVs are physically connected by a pulling rope to ensure that the platoon maintains the same travelling direction. 
The ARV is built based on a low-cost robot WIFIBOT~\cite{wifibot2018}. Mechanical design and four wheel drive of WIFIBOT allow the ARV to move over irregular surfaces or even small obstacles. Moreover, the small dimensions (length = 28 cm, width = 30 cm, and height = 20 cm) and low weight of 4.5 kg make the ARV easily transportable and manageable during the experiments. 

With regards to the wireless communication interface, the Crossbow TelosB wireless node mounted on a 1m-high plastic pole is placed on top of the ARV. Specifically, the TelosB node is a low power wireless sensor module equipped with an IEEE 812.15.4-compliant RF transceiver (the Chipcon CC2420 operating in the 2.4 GHz frequency band), a built-in antenna, and an 8 MHz TI MSP430 microcontroller. The TelosB node has the maximum data rate of 250 kbps, while the maximum transmission power is 0 dBm. In particular, the data rate and transmission power of all ARVs are set to the maximum level during our experiments. In terms of packet length, the payload of the data packet has 100 bytes while the beacon packet is 1 byte. Although the TelosB node is designed for low data rate transmission and low computation capabilities, it is still applicable for executing CoopKey at the ARV testbed due to a short data packet length. 
Moreover, we also connect the TelosB node at the tail ARV, i.e., $v_4$, to a laptop via a USB connection to record the secret key and data packet at the ARV for postprocessing and analysis. 

\begin{figure}[htb]
\centering
\includegraphics[width=4in,height=1.5in]{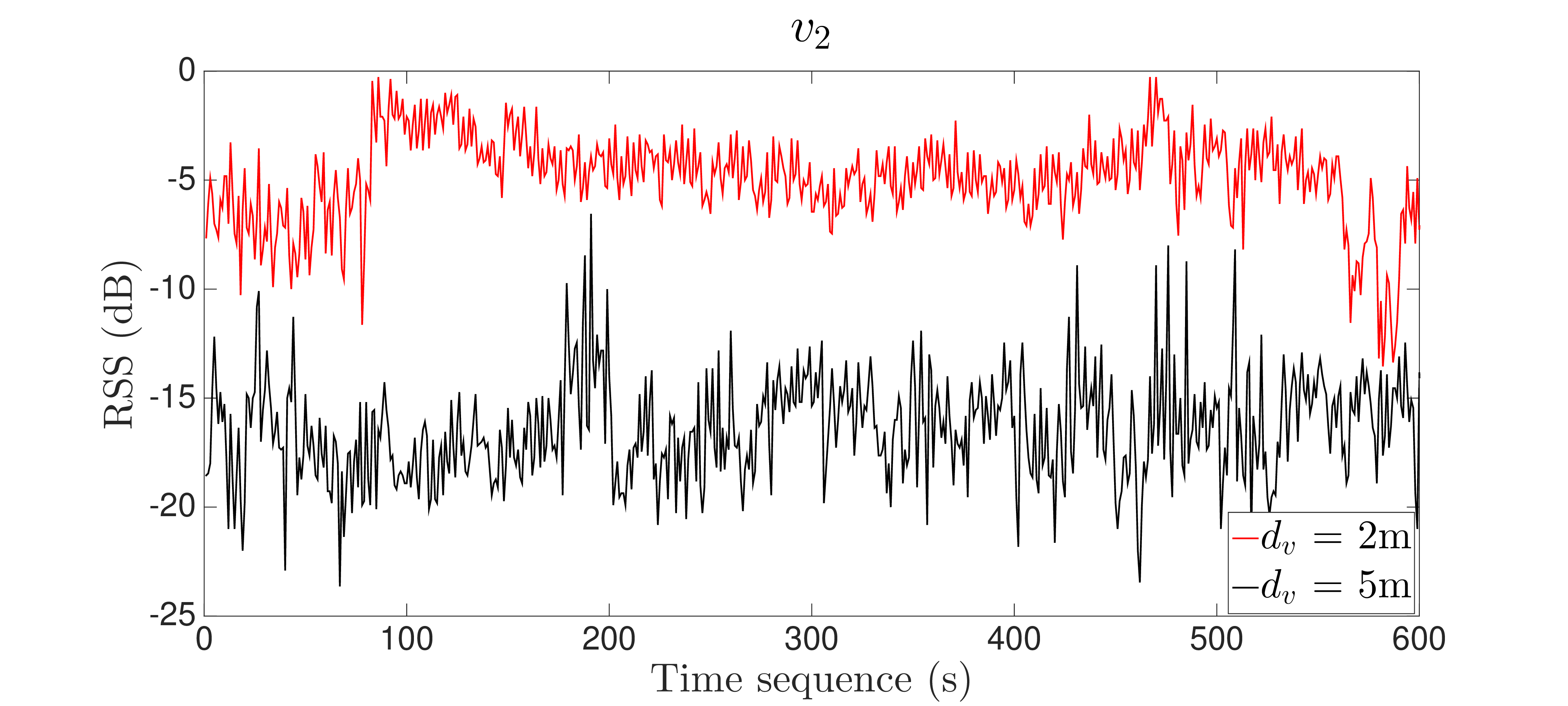} \\
\includegraphics[width=4in,height=1.5in]{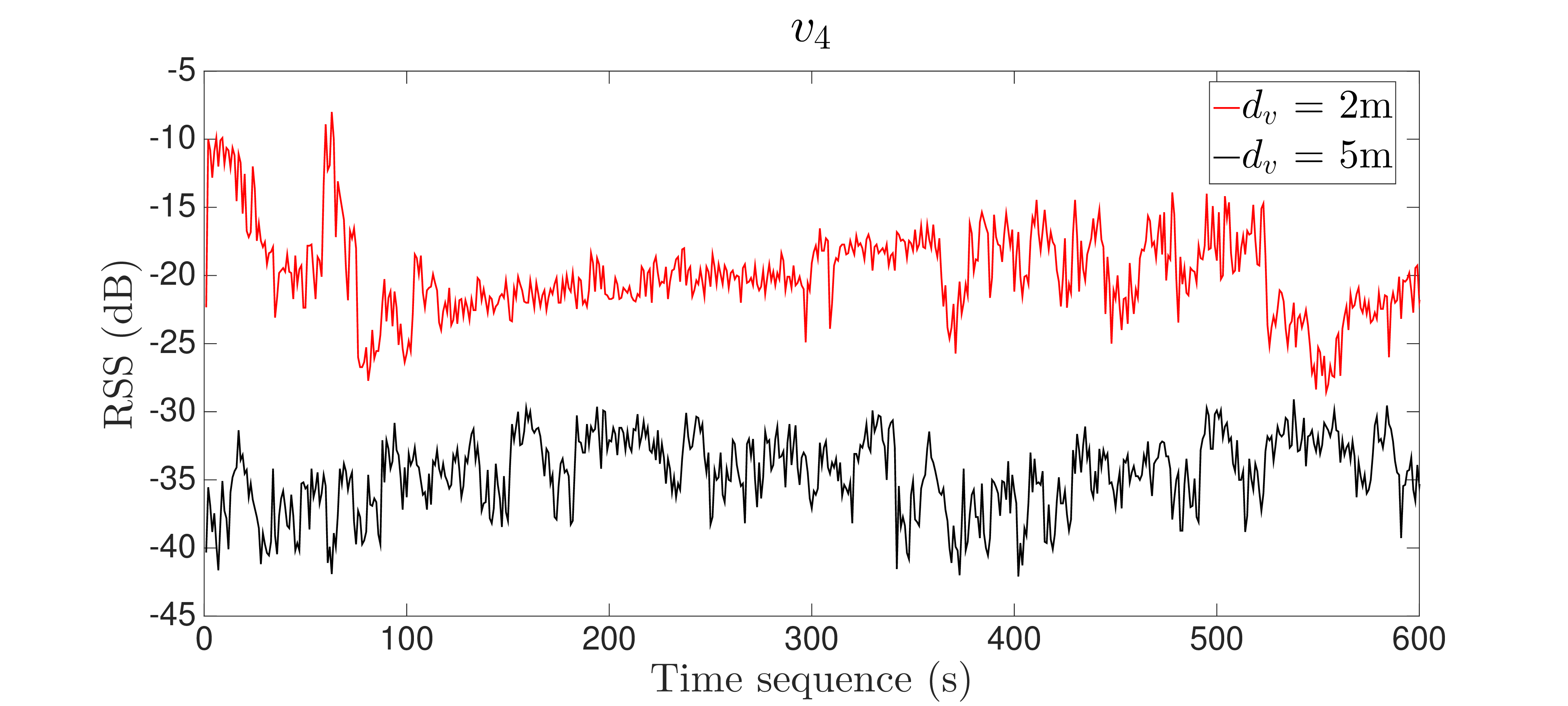}
\caption{RSS of the first and the last following ARVs with regards to 2 m or 5 m of the inter-ARV distance (i.e., $d_v$).} 
\label{fig_rssi}
\end{figure}

The transmission of data packets is initialized by the lead ARV. The data packets are encrypted by CoopKey at the lead ARV, and immediately disseminated to its adjacent following ARV all the way to the tail ARV. When the tail ARV successfully receives the data, it broadcasts an acknowledgement packet so that the lead ARV can transmit a new packet. In case of packet loss during the dissemination, a timeout of the packet dissemination at the lead ARV is set to 3 seconds. In other words, the lead ARV disseminates a new data packet if the acknowledgement from the tail ARV is not received within 3 seconds. 
Moreover, an experiment is conducted on the ARV testbed to measure RSS at the three following ARVs with the different inter-ARV distance. Figure~\ref{fig_rssi} shows that RSS at the following ARVs drops with the inter-ARV distance, which demonstrates feasibility of the channel estimation in CoopKey. 

\section{Experimental Results}
\label{sec_exp}
In this section, we first present experimental scenarios and performance metrics for evaluating CoopKey. Then, extensive experiments are conducted on the ARV testbed to show the Bit MisMatch Rate (BMMR) of PVCPS with regards to inter-ARV distances, RSS quantization intervals, and $Z$ iterations. 
To further reveal the security of CoopKey, we show BMMR of the secret key that is generated by the eavesdropper, when it overhears the beacon packets. 
In addition, to study the effect of system scalability on the performance, CoopKey is also evaluated in simulations with an extended platoon size. 

\subsection{Experimental setup}
The ARVs in our testbed travel in a straight line with the velocity about 0.3 $\sim$ 1 m/s which is determined by the lead ARV, keeping the operations of the cruise control simple. 
Although the experiments are conducted at a low speed, the performance evaluation is still convincing since the proposed secret key agreement is achieved based on the optimal quantization of the RSS no matter what speed the PVCPS drives. 
Duration of one experiment (i.e., traveling time of the ARVs) is around 10 minutes. The inter-ARV distance increases from 2 m to 8 m given $P^{tx}_i(t) = 0$ dBm. The number of RSS quantization intervals, i.e., $L$, is 2 or 5. To explore the impact of $Z$, CoopKey is conducted in 1, 5, 10, 15, or 20 iterations. 
Moreover, payload of the beacon packet contains PacketType and SenderID. The field PacketType is set to ``1'' for the beacon packet, and ``0'' for all other data types. The ACK packet of the tail vehicle has one bit, where ``1'' indicates that the data is successfully received, otherwise, it is ``0''.

We also carry out a comparison study between CoopKey and the non-cooperative key generation scheme (named as ``LocalKey''), where each following ARV separately generates its secret key based on the quantized RSS measurement when the beacon packet is received. In terms of the performance metric, BMMR of PVCPS defines the number of secret bits generated by the following ARV, which mismatches the one generated by the lead ARV, over the $Q$ bits. Thus, it gives 
\begin{align}
\text{BMMR} = \frac{1}{Q} \sum^Q_{q=1} \veebar(K_1(q), K_i(q)), 
\label{eq_def_bmmr}
\end{align}
where $K_i(q)~(i \neq 1)$ is the $q$-th bit of the key generated by $i$-th ARV. A mobile device is considered as the eavesdropper which is 2 meters away from the second ARV in our testbed. The eavesdropper travels in parallel to the platooning ARVs with the similar velocity. A TelosB node is placed on the eavesdropper for overhearing the transmission of the platooning ARVs. Moreover, the eavesdropper also applies CoopKey to generate its secret key for decoding the overheard data packets. 

\subsection{Performance of secret key agreement}
\subsubsection{BMMR}
\begin{figure}[htb]
\centering
\includegraphics[width=5in]{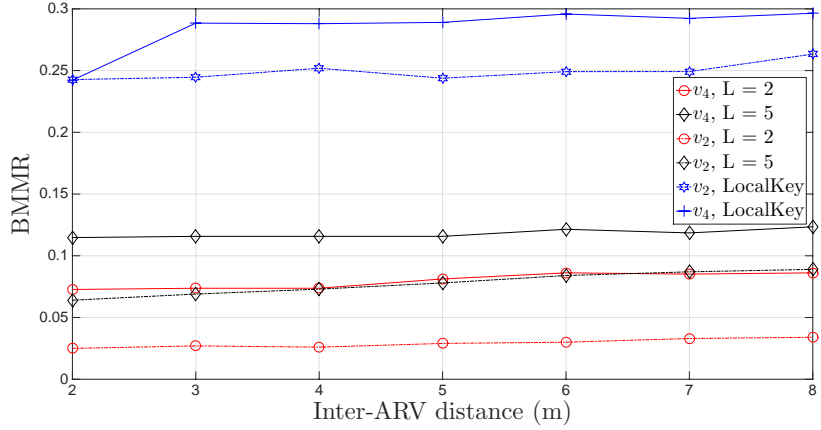}
\caption{BMMR of CoopKey at the ARVs $v_2$ and $v_4$ with an increasing inter-ARV distance $d_v$, given $L$ = 2 or 5.} 
\label{fig_bmmr_distance}
\end{figure}

Figure~\ref{fig_bmmr_distance} shows BMMR at $v_2$ and $v_4$ with an increasing inter-ARV distance, where $Z$ = 1 and $L$ = 2 or 5. We can see that CoopKey with $L$ = 2 achieves about 22\% lower BMMR than LocalKey at $v_2$ and $v_4$. This is because CoopKey recursively adapts the quantization intervals at each ARV based on the measured/estimated RSS readings for generating a unanimous secret key. Moreover, BMMR of CoopKey at $v_2$ and $v_4$ gradually increases with the inter-ARV distance due to channel estimation errors caused by the RSS measurement randomness. Particularly, BMMR of CoopKey at $v_2$ is lower than the one at $v_4$ by 4\% when $d_v$ = 2 m since $v_2$ generates the secret key with the RSS readings of the beacon packet, while $v_4$ generates the key with the estimated RSS of $H_{1,2}(t)$. 

We also observe in Figure~\ref{fig_bmmr_distance} that decreasing the quantization intervals, i.e., $L$, reduces BMMR of CoopKey. Specifically, BMMR at $v_2$ and $v_4$ drops from 0.06 to 0.02, and from 0.12 to 0.07, respectively, while $L$ decreases from 5 to 2. This indicates that the small number of quantization intervals makes the vehicles easy to reach key agreement. However, decreasing $L$ results in a rise of security vulnerability on the generated secret key, where the eavesdropper may generate the same secret key as the platooning vehicles. Therefore, it is critical to comprehensively configure $L$ according to the required BMMR and the platoon size.

\begin{figure}[htb]
\centering
\includegraphics[width=4.5in]{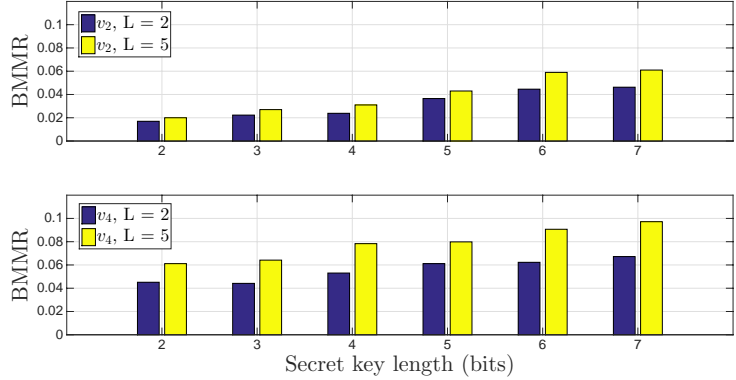}
\caption{BMMR of CoopKey at the ARVs $v_2$ and $v_4$ with different key length $Q$, given $L$ = 2 or 5.} 
\label{fig_bmmr_keylength}
\end{figure}

Figure~\ref{fig_bmmr_keylength} demonstrates BMMR at $v_2$ and $v_4$ with the growth of secret key length $Q$, where $Z$ = 1, $d_v$ = 2 m, and $L$ = 2 or 5. Generally, BMMR of CoopKey increases with the secret key length. The reason is because the longer the secret key is, the more secret bits need to be generated and unified, and in turn, the lower chance that the quantized RSS readings become consistent. In particular, CoopKey with $Q$ = 2 at $v_2$ achieves the minimum BMMR which is about 0.019 (at $L$ = 2) or 0.02 (at $L$ = 5). Moreover, when the key length increases to 7 bits, BMMR at $v_2$ is less than 0.061. At $v_4$, the BMMR of CoopKey with $L$ = 2 and 5 is smaller than 0.065 and 0.1 when $Q$ is less than 7 bits. 

Figure~\ref{fig_bmmr_keylength} implies a tradeoff between BMMR of CoopKey and command dissemination security. Shortening the secret key length of CoopKey reduces BMMR, however, a drop in the number of secret bits results in a fall of transmission security, where the eavesdropper may generate the same key to decode the data. Therefore, it is critical to holistically configure $Q$ in PVCPS according to the minimum requirement of the BMMR and quantization intervals. 

\begin{figure}[htb]
\centering
\includegraphics[width=4.5in]{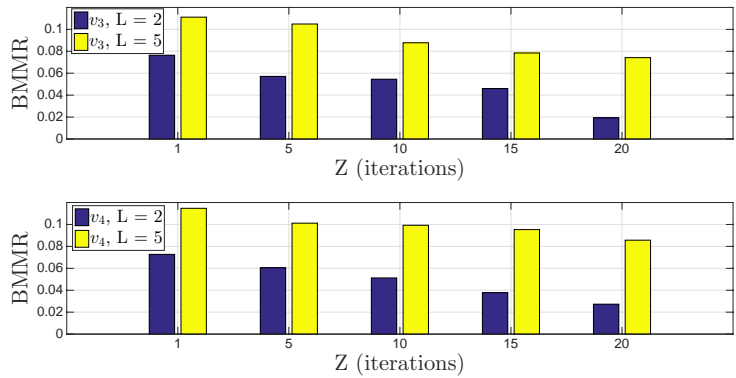}
\caption{BMMR of CoopKey at the ARVs $v_2$ and $v_4$ with an increasing $Z$, given $L$ = 2 or 5.} 
\label{fig_bmmr_iterations}
\end{figure}

In Figure~\ref{fig_bmmr_iterations}, $Z$ of CoopKey increases from 1 to 20 iterations, given $L$ = 2 or 5 and $d_v$ = 2 m. In this case, CoopKey is conducted in $Z$ iterations ($Z \geq$ 1), and $Z$ number of secret keys are generated at each ARV. It is observed that BMMR generally decreases with the growth of $Z$ values. In particular, given $L$ = 2, the BMMR at $v_3$ and $v_4$ drops from 0.076 and 0.07 to 0.02 and 0.025, respectively. It confirms that the larger $Z$ is, the smaller the estimation errors are, and in turn, the higher likelihood that the keys become consistent due to $1 - (1/R_{BMM})^Z$. Therefore, increasing $Z$ can reduce BMMR of CoopKey. 

\subsubsection{Secret bits randomness}
\begin{table}[htb]
\centering
\caption {P-values from NIST statistical test suite, where $d_v = $ 2, 3, 4, 5, 6, 7, or 8. To pass the test, all p-values must be greater than 0.01.}
\begin{tabular}{|l|p{1cm}|p{1cm}|p{1cm}|p{1cm}|p{1cm}|p{1cm}|p{1cm}|}
    \hline
    Test                  & A & B & C & D & E & F & G \\ \hline
    Frequency             & 0.21 & 0.53 & 0.12 & 0.02 & 0.91 & 0.07 & 0.53 \\ \hline
    Block Frequency       & 0.74 & 0.534 & 0.35 & 0.53 & 0.74 & 0.35 & 0.21 \\ \hline
    Cumulative sums(Fwd)  & 0.21 & 0.35 & 0.21 & 0.04 & 0.53 & 0.74 & 0.74 \\ \hline
    Cumulative sums (Rev) & 0.91 & 0.07 & 0.12 & 0.04 & 0.54 & 0.21 & 0.53 \\ \hline
    Runs                  & 0.21 & 0.74 & 0.53 & 0.21 & 0.79 & 0.74 & 0.91 \\ \hline
    longest run of ones   & 0.12 & 0.35 & 0.74 & 0.53 & 0.91 & 0.12 & 0.07 \\ \hline
    FFT                   & 0.12 & 0.74 & 0.02 & 0.07 & 0.12 & 0.02 & 0.07 \\ \hline
    Approx. Entropy       & 0.21 & 0.35 & 0.35 & 0.91 & 0.74 & 0.07 & 0.35 \\ \hline
    Serial                & 0.21, 0.07 & 0.74, 0.35 & 0.35, 0.53 & 0.12, 0.99 & 0.35, 0.91 & 0.35, 0.02 & 0.74, 0.53 \\ \hline
\end{tabular}
\label{table_nist}
\end{table}

To ensure that the secret key generated is substantially random, the standard randomness test suite from NIST~\cite{rukhin2001statistical} is employed to verify the effectiveness of CoopKey. Given 16 different statistical tests in the NIST test suite, we run 8 tests of them and calculate their p-values. The p-value indicates the probability that a perfect random number generator would have produced a sequence less random than the input sequence that is tested. The reason of selecting the 8 NIST tests is because their recommended input size meets bit streams of the secret keys in our experiments. Note that the remaining 8 tests require a very large input bit stream (more than $10^6$ bits), where a large number of keys (in gigabytes) need to be generated. Moreover, each test is conducted in 7 scenarios from A to G, where $d_v$ increases from 2 m to 8 m. 

As shown in Table~\ref{table_nist}, all the keys generated by CoopKey pass the test, and have much larger p-value than 0.01 which is the threshold to pass the test. In particular, a p-value larger than 0.01 indicates that the secret bit streams of CoopKey are random with a confidence of 99\%. Furthermore, the randomness of the keys generated by CoopKey substantially increases the time complexity of cracking the keys at the eavesdropper, hence protecting the data dissemination from the eavesdropping attacks. 

\subsection{BMMR of the eavesdropper}
\begin{figure}[htb]
\begin{center}
\includegraphics[width=3.5in]{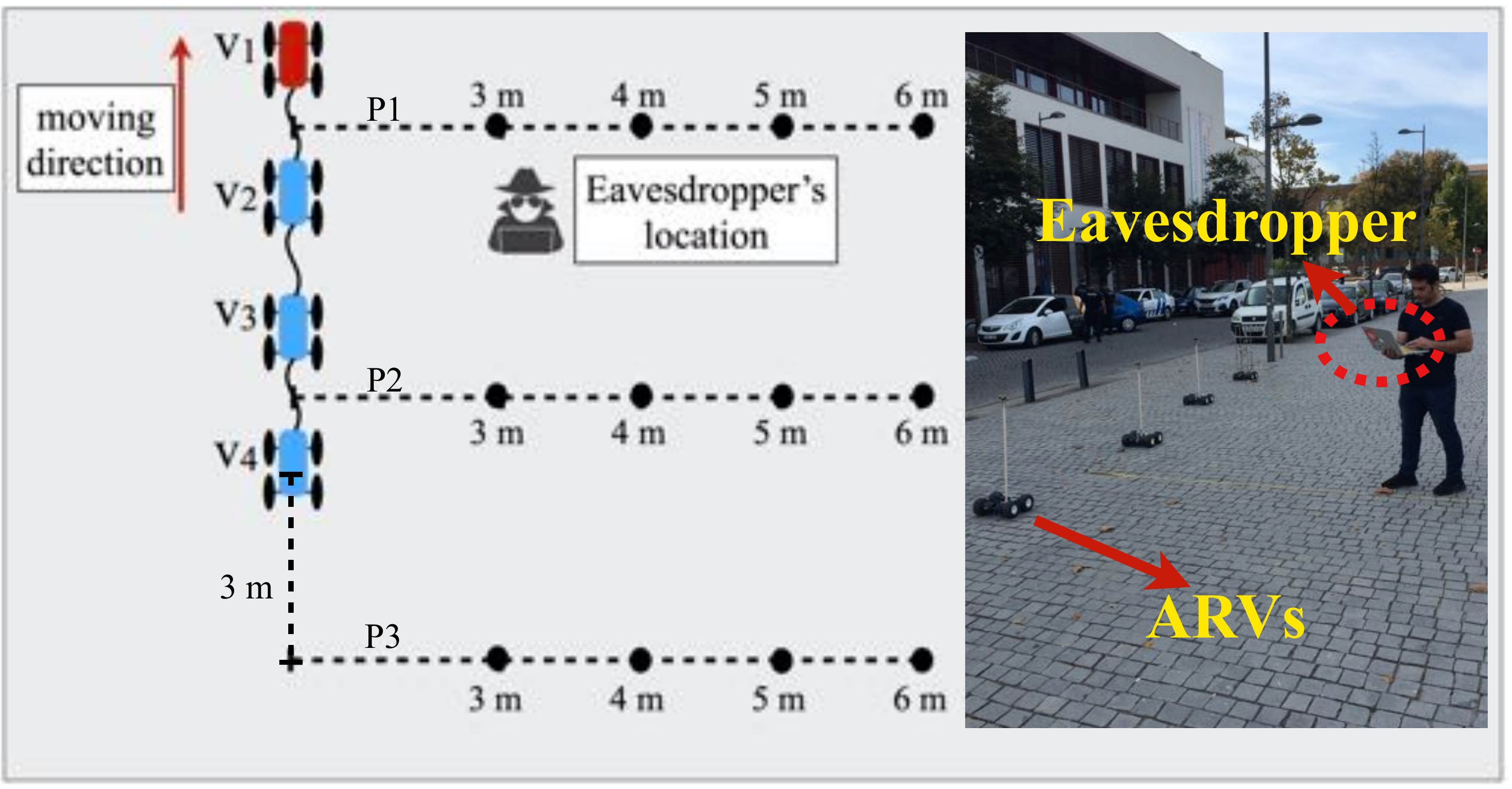} \\
\includegraphics[width=4in]{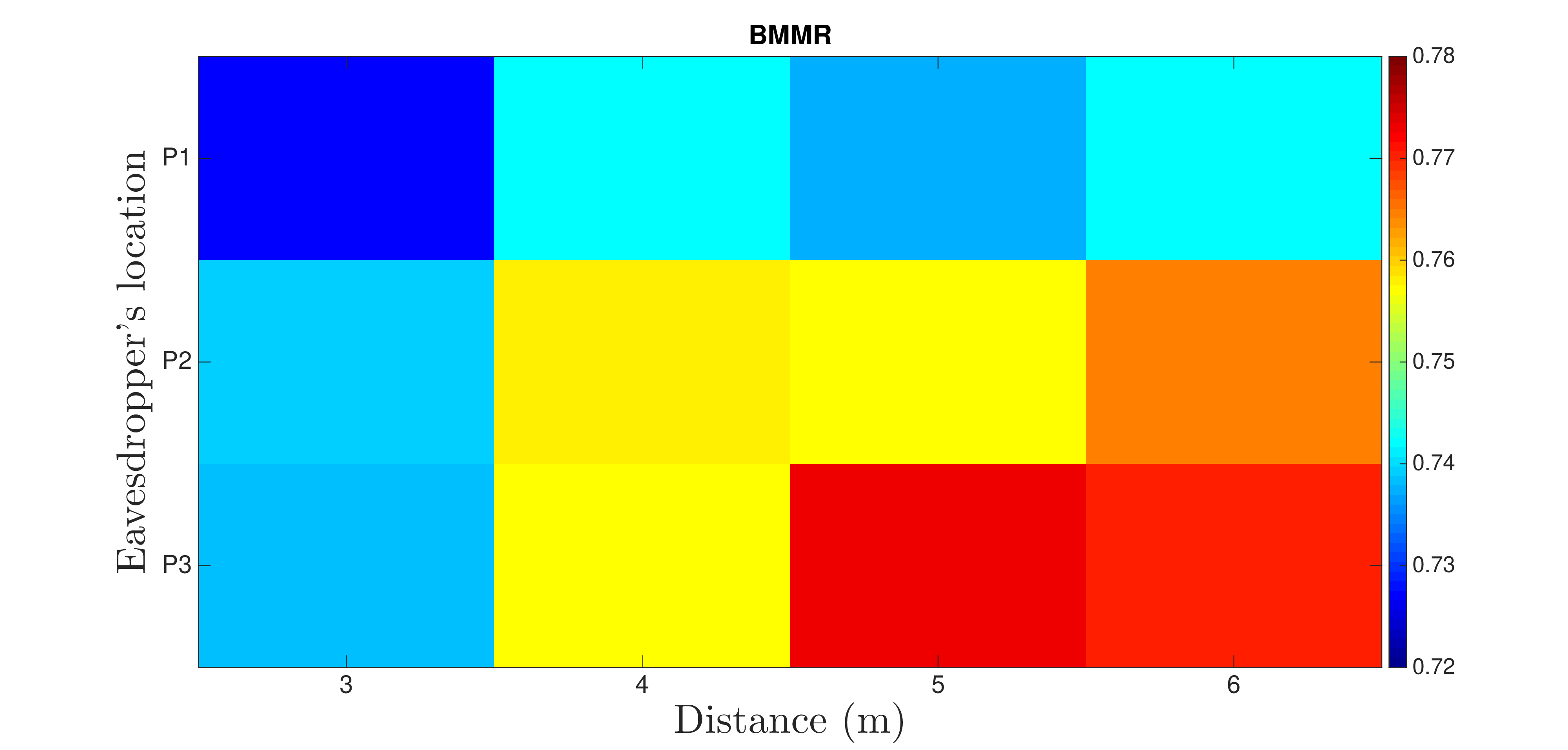} 	
\end{center}
\caption{BMMR of the eavesdropper with regards to its relative locations to the platoon.}
\label{fig_eavesdropper}
\end{figure}

To further unveil the security of CoopKey, Figure~\ref{fig_eavesdropper} plots the BMMR of the eavesdropper with regards to its relative locations to the platoon. The BMMR of the eavesdropper calculates the number of secret key bits generated by the eavesdropper, which mismatch the key bits generated by the platooning vehicle. Therefore, the higher BMMR the eavesdropper's decoded data has, the more secure key agreement CoopKey achieves. 

We consider three specific locations of the eavesdropper, i.e., P1, P2, and P3. P1 is the location of the eavesdropper next to the middle of $v_1$ and $v_2$, P2 is the one next to the middle of $v_3$ and $v_4$, and P3 is the one 3 m behind $v_4$. For each location, the distance between the eavesdropper and the platoon enlarges from 3 m to 6 m. Here, we assume that the eavesdropper can be identified when the distance is less than 3 m.

As observed, the lowest BMMR is about 0.73, where the eavesdropper is 3 m away from the platoon at P1. With the growth of the distance, the BMMR of the eavesdropper increases to 0.75. Furthermore, when the eavesdropper is 6 m away from the platoon at P3, its BMMR is about 0.77. The reason is that randomness of the RSS measurements at the eavesdropper is much higher than the one at the platooning ARVs due to misalignment with the ARVs. As a result, the quantization intervals of the eavesdropper could be very different from the ones used by the ARVs given~\eqref{eq_def_Delta}. A high BMMR at the eavesdropper indicates that the eavesdropper is not able to recover the secret key generated by CoopKey at the ARVs, even though the eavesdropper has the knowledge of CoopKey. The key agreement achieved by CoopKey is highly secure against the eavesdropping attack.

\subsection{Runtime measurement}
In this experiment, the disseminated data packets from the lead ARV to the tail one are encrypted by CoopKey (where $Z$ = 1, 5, 10, 15, or 20). Table~\ref{tb_runtime} shows average end-to-end latency of the data dissemination, where we repeat the experiment for ten times at each setting of $Z$. Particularly, the runtime is calculated by summing up the execution time of CoopKey, the data transmission time at the ARVs, and the propagation delay of the data packet. From Table~\ref{tb_runtime}, we can see that the average latency grows with the increase of $Z$ iterations. This is reasonable because Algorithm~\ref{alg_dpa} is conducted to derive the optimal quantization intervals with multiple iterations, which results in extra execution time. 

\begin{table}[htb]
    \centering
    \caption{Runtime measurement of CoopKey}
    \begin{tabular}{|c|c|c|c|c|c|c|c|c|c|c|c|}
    \hline
    \multirow{2}{*}{$Z$ iterations} & \multicolumn{10}{c|}{End-to-end dissemination latency (ms)} & \multirow{2}{*}{Average latency (ms)} \\ \cline{2-11}
		& 1  & 2  & 3  & 4  & 5  & 6  & 7  & 8  & 9  & 10 &  \\ \hline
	1       & 81 & 76 & 80 & 82 & 87 & 81 & 66 & 93 & 97 & 87 & 83 \\ \hline
	5       & 101 & 93 & 93 & 82 & 96 & 94 & 85 & 98 & 92 & 99 & 93.3 \\ \hline
	10     & 104 & 98 & 100 & 97 & 100 & 108 & 101 & 105 & 102 & 103 & 101.8 \\ \hline
	15     & 100 & 108 & 103 & 107 & 107 & 160 & 100 & 103 & 469 & 104 & 146.1 \\ \hline
	20     & 620 & 774 & 619 & 780 & 597 & 515 & 777 & 613 & 620 & 608 & 652.3 \\ \hline
\end{tabular}
\label{tb_runtime}
\end{table}

Based on Figure~\ref{fig_bmmr_iterations} and Table~\ref{tb_runtime}, a tradeoff between the BMMR of CoopKey and the data dissemination latency can be known. Specifically, the higher $Z$ leads to the more unified secret key bits in CoopKey while sacrificing timeliness of the data dissemination. Therefore, the parameter $Z$ has to be chosen to reduce the BMMR of CoopKey for the key agreement while meeting the critical need for the low-latency data dissemination in PVCPS.

\subsection{Scalability study}
\begin{figure}[htb]
\centering
\includegraphics[width=5in]{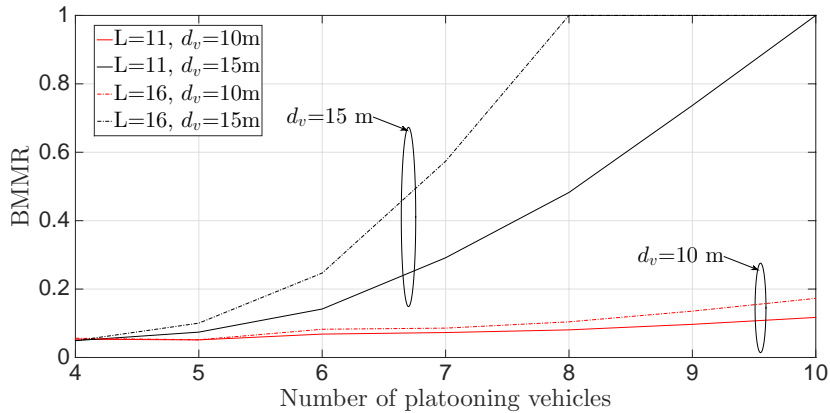}
\caption{BMMR of CoopKey with respect to the platoon size $N$, inter-vehicle distance $d_v$, and number of RSS quantization intervals $L$.} 
\label{fig_agreeCars}
\end{figure}

To study the effect of system scalability on the performance, CoopKey is evaluated with an extended platoon size based on simulations. Figure~\ref{fig_agreeCars} plots the BMMR of CoopKey in terms of the platoon size of PVCPS, where $L$ is set to 11 or 16. The distance between the two vehicles is maintained at $d_v$ which is set to 10 or 15 meters. 

The BMMR of CoopKey increases with the growth of $N$. Specifically, when $N = 4$, the CoopKey schemes with $d_v = 10$ m and $d_v = 15$ m have a similar BMMR. Moreover, when $N$ $\geq$ 5 vehicles, CoopKey with $d_v = 10$ m has lower BMMR than the one with $d_v = 15$ m. In particular, the BMMR of CoopKey with $d_v = 15$ m rises from 6\% to 100\%, while the one with $d_v = 10$ m increases less than 15\%. This confirms the fact that a short inter-vehicle distance in~\eqref{eq_distance} leads to a strong RSS which effectively reduces BMMR. 

In Figure~\ref{fig_agreeCars}, it is also observed that increasing the quantization intervals, i.e., $L$, results in a high BMMR. For example, in the case that $d_v = 15$ m and $d_v = 10$ m, CoopKey with $L = 16$ has 50\% and 8\% more BMMR than the one with $L = 11$, when $N = 8$ and $10$, respectively. 

\section{Literature Review: Secret Key Generation In Mobile Networks}
\label{sec_relatedwork}
In this section, we review the literature on secret key extraction from the RSS variations in mobile networks. 

\subsection{Link-based secret key generation}
The effectiveness of link-based secret key extraction between two wireless devices, e.g., temporal-spatial variations in the radio channel, and mobility of the devices, is experimentally measured~\cite{jana2009effectiveness}. An environment adaptive secret key generation scheme is developed to improve secret bit generation rate by extracting multiple bits from each RSS measurement. 
In~\cite{wan2018physical}, a secured communication scheme is studied for automotive wireless communication. Symmetric cryptographic keys are generated between two vehicles, based on physical randomness of the automotive wireless channel under memory and performance budgets. 
A secret key generation framework, called HRUBE, extracts the secret bits from a series of radio channel measurements between two wireless devices~\cite{patwari2010high}. The channel measurement is quantized by HRUBE to an arbitrary number of secret bits without censoring. However, HRUBE requires to precisely know apriori statistical knowledge of the channel distribution to obtain the key length, which is not practical in real-world environments. To address this limitation, Croft \textit{et al.} study a ranking method to remove non-reciprocities of the unknown channel characteristics between the two devices~\cite{croft2010robust}. The ranking method also enables the secret bit extraction process independent of the unknown channel distribution. 

For the key agreement of two mobile devices, the impact of mobility patterns in obtaining the uncorrelated channel measurements is studied in~\cite{zhang2010mobility}. It is found that channel impulse responses are mostly uncorrelated when movement step size is larger than one foot. Moreover, the measured channel impulse responses are encoded, and the mismatched secret bits between the two devices can be reconciled by using forward error correction. 

The channel response from multiple Orthogonal Frequency-Division Multiplexing (OFDM) subcarriers can provide the channel information for the RSS-based key generation~\cite{liu2013fast}. A channel gain complement method is developed to reduce the non-reciprocity of RSS in the key generation. In~\cite{epiphaniou2018nonreciprocity}, the vehicle-to-vehicle communication characteristics, e.g., multipath propagation and surrounding scatterers' mobility, are incorporated in the key generation process. The non-reciprocity compensation method in~\cite{liu2013fast} is used by the transmitter to generate the secret key according to designated RSS, while Turbo codes are used for channel information reconciliation of multiple links. A stochastic vehicular channel model is utilized to generate the receiver's channel response. 

However, most works in the literature are interested in generating the key to encrypt point-to-point communications based on mutually-known channel information. This can hardly meet the critical need for the key agreement of multiple users, e.g., vehicular platoon, where the secret key has to be generated and conformed at each vehicle based on the local channel observation.

In~\cite{liu2012collaborative}, relay nodes are deployed to assist the RSS-based key generation between two devices. The relay nodes send the difference of RSS over different radio channels to the two devices. A framework is developed for the two devices to utilize the information received from the relay nodes and RSS measurements between the relay nodes and themselves to generate the secret key. 
Wang \textit{et al.} present a key generation protocol in narrowband fading channels, where the sender and the receiver extract the channel randomness with the aid of relay nodes~\cite{wang2012cooperative}. Their protocol applies a time-slotted key generation scheme, where each relay node contributes a small portion of key bits so that the complete global key bit information is not available to the relays.

However, the key generation with the aid of relay nodes is not applicable to PVCPS since employing relay vehicles can be costly. In addition, the key agreement with relays would cause extra latency on the control command dissemination, which can lead to cruise control failures due to lack of timely updates on the driving status.

\subsection{Application-dependent secret key generation}
Social ties of mobile devices, which characterize the strengths of relationships among mobile users, can be leveraged to generate the secret key with the assistance of relay pairs~\cite{waqas2018social}. On the basis of social ties, the selection of relay pairs is formulated by coalition game theory for improving secure key generation rate while protecting the keys secret from both eavesdropper and non-trusted relays. 
In~\cite{jiang2019shake}, a handshake-based pairing scheme between wrist worn smart devices is developed based on the observation that, by shaking hands, both wrist worn smart devices conduct similar movement patterns. A device pairing scheme is developed by exploiting the motion signal of the devices generated by the handshake to negotiate a secret key between users. A fuzzy cryptography algorithm is further studied to remove distortion of the extracted acceleration data, thus ensuring the robustness of the key agreement. 
In~\cite{lin2019h2b} and~\cite{pirbhulal2018heartbeats}, heartbeat intervals measured by wearable medical devices are used as a random source to generate secret keys. The heartbeat intervals can be sampled by electrocardiogram sensors or piezo vibration sensors. It is shown that the heartbeat-based secret key is secured against typical attacks and power-efficient in wireless body sensor networks.

A random secret key generation scheme that integrates differential logical pattern method is presented in~\cite{thangamani2019lightweight}. In particular, an input message is split into a number of blocks for pattern extraction. A random key is generated based on the generated pattern, where the differential logical pattern method performs the encryption process with differential mode of the input message. 
A key agreement protocol is studied for user authentication in Internet of Things (IoT) networks~\cite{wazid2017design}. The key generation applies cryptographic hash function along with the symmetric encryption/decryption, which supports various functionality features, such as user login, sensing node registration, and biometric update. 
Biometric information can be used for key generation and agreement between two parties over an open network~\cite{al2012robust}. To improve robustness of the key agreement, random orthonormal projection and biometric key binding are explored to combine biometrics with existing authentication factors. 

\section{Conclusions}
\label{sec_cond}
In this paper, we study the CoopKey scheme for securing the vehicular control command dissemination in PVCPS. The secret key is generated based on the quantized RSS measurements of the inter-vehicle radio channel. The RSS quantization intervals are recursively adjusted until a unanimous secret key is generated for encrypting/decrypting the disseminated command. For evaluating performance of CoopKey, a platooning testbed is built on multiple ARVs and TelosB wireless nodes. Extensive experiments demonstrate that CoopKey achieves significantly lower secret bit mismatch rate with respect to the platoon size, the inter-vehicle distance, and the number of quantization intervals. CoopKey is also evaluated in simulations with an extended platoon size and inter-vehicle distance to study the effect of system scalability on the performance. 

%
\section{Acknowledgement}
This work was supported by National Funds through FCT/MEC (Portuguese Foundation for Science and Technology) and co-financed by ERDF (European Regional Development Fund) under the PT2020 Partnership, within the CISTER Research Unit (CEC/04234); also by FCT/MEC and the EU ECSEL JU under the H2020 Framework Programme, within project ECSEL/0002/2015, JU grant nr. 692529-2 (SAFECOP).

The authors would like to thank Mr. Marwin Adorni for his assistance with the testbed setup. The authors also thank the editors and the anonymous reviewers for their constructive comments on the article.

\bibliographystyle{elsarticle-num}
\bibliography{coopkey}

\end{document}